\documentclass[pra,preprint]{revtex4} 
\usepackage{graphicx,color} 
\usepackage{bm}       
\usepackage{amsmath}  
\usepackage{amssymb}

\newcommand{\KHS}{\mathcal{K}}
\newcommand{\kt}[1]{\ensuremath{|{#1}\rangle}}
\newcommand{\br}[1]{\ensuremath{\langle {#1}|}}

\begin{document}
\title{Five is More: Comments on Symmetry, Integrability, and Solvability for a Few Particles in a One-Dimensional Trap}

\author{N.L.~Harshman\footnote{Electronic address: harshman@american.edu}}

\affiliation{Department of Physics\\
4400 Massachusetts Ave. NW\\ American University\\ Washington, DC 20016-8058}

\begin{abstract}

This contributed conference proceeding reviews some results about a system of a few identical particles with spin trapped in one-dimensional potentials and experiencing two-body interactions. The focus is on how symmetry, integrability, and solvability depend on the trap shape, two-body interaction, the number of particles, and the number of spin components. A series of comments are presented that characterize the minimal symmetries possible for a composite system constructed from interacting single particles, with special focus on the contact interaction. For five and more particles with internal components like spin, a kind of universality called algebraically solvability is lost.
\end{abstract}%
\maketitle
\section{Introduction}
\label{intro}

This brief conference proceedings considers the symmetries of a model of $N$ particles with $J$ spin components in one-dimensional trap. This model has attracted much interest recently because many-body~\cite{manybody} and few-body~\cite{fewbody} experiments with ultracold atoms in highly-elongated traps and interacting via tunable Feshbach resonances are well-modeled as particles in one-dimension with contact interactions~\cite{oneD}. This motivates examining the symmetries and solvability of the following $N$-body Hamiltonian:
\begin{equation}\label{hammy}
\hat{H} = \hat{H}_0 + \sum_{\langle i ,j \rangle} \hat{V}_{ij}
\end{equation}
where the non-interacting part is the sum of one-particle Hamiltonians
\begin{equation}
\hat{H}_0 = \sum_{i=1}^N \hat{H}^1_i = \sum_{i=1}^N \left(\frac{1}{2m} \hat{P}_i^2 + V^1(\hat{X}_i) \right)
\end{equation}
and the two-body interactions are spin-independent functions of interparticle separation $\hat{V}_{ij} = V^2(|\hat{X}_i - \hat{X}_j|)$.

This model Hamiltonian also has a long history as a touchstone of theoretical and mathematical physics applied to few-body and many-body systems~\cite{longhist}. Depending on the shape of the trap $V^1$, the interaction potential $V^2$, the number of particles $N$, and number of internal particle components $J$, the system may be integrable, analytically solvable, or even algebraically solvable~\cite{solve}. These properties have intrinsic interest to mathematical physicists, and they also have ramifications for physical properties like thermalization~\cite{thermal}, universality~\cite{universality}, and entanglement~\cite{entanglement}, and practical matters like precision calculation and experimental control (c.f.\ recent reviews~\cite{reviews}).

The following comments are intended to provide a cross-section of results about the symmetries of the Hamiltonian. Some of these results are old and familiar, but some are from the recent wave of analysis inspired by current experiments. Others are new results presented at the conference for the first time and culled from my recent two-part article ``One-Dimensional Traps, Two-Body Interactions, Few-Body Symmetries''~\cite{Harshman2015a,Harshman2015b}. For a detailed set of references to classic and recent work, and for a systematic development of the results summarized in these comments, see those two articles.

\section{Eight Comments}

\begin{enumerate}

\item One-particle systems in one dimension are Liouville integrable.

For any trap potential $V^1$, the one-particle Hamiltonian $H^1$ is in involution with itself, and so there are as many invariants as degrees of freedom, namely, one!  Classically this is enough to solve for the equation of motion; quantum mechanically it means that a single observable $H^1$ and its spectrum is enough to completely separate the spatial Hilbert space $\KHS$  into subspaces characterized by energy. In addition to being integrable, some potentials are solvable, and some are even algebraically solvable. For truly trapped systems, the spectrum $\sigma_1$ is discrete, countably-infinite, and non-degenerate, i.e.\ $\sigma_1 = \{\epsilon_0, \epsilon_1, \epsilon_2, \cdots\}$. The associated kinematic symmetry $\mathrm{K}_1$ is time translation $\mathrm{T}_t$ represented on the spatial Hilbert space by the exponentiation of the Hamiltonian. If the trap is parity symmetric, then the kinematic symmetry is isomorphic to $\mathrm{K}_1 \sim \mathrm{O}(1) \times \mathrm{T}_t$ and there is another integral of the motion, the parity operator $\hat{\Pi}$. For harmonic traps, the one-particle kinematic symmetry is $\mathrm{K}_1 \sim \mathrm{U}(1) \times \mathrm{T}_t$ because of the rotational invariance of phase space for a harmonic oscillator.

\item The Hamiltonian $\hat{H}^N$ has at least the minimal kinematic symmetry $\mathrm{K}_N \supseteq \mathrm{P}_N \times \mathrm{K}_1$.

Particle permutation symmetry $\mathrm{P}_N \sim \mathrm{S}_N$ is clear from the Hamiltonian and the factor of $\mathrm{K}_1$ is a consequence of the Galilean-invariance of the two-body interactions. This minimal symmetry is not enough to integrate the problem for general traps and interactions, but it does imply that (unless there are emergent symmetries or accidental degeneracies) every energy level should have a degeneracy corresponding to the dimension of an irreducible representation of $\mathrm{S}_N$. In the case of a harmonic traps, both total parity and relative parity are invariants, and so there is an extra factor of  $\mathrm{O}(1)$ in $\mathrm{K}_N \sim \mathrm{S}_N \times \mathrm{O}(1) \times  \mathrm{U}(1) \times \mathrm{T}_t$. This is enough extra symmetry for the case with $N=2$ to be integrable for any $\hat{V}_{12}$, and solvable for contact interactions~\cite{avakian}.

\item The non-interacting Hamiltonian $\hat{H}^N_0$ has at least the minimal kinematic symmetry  $\mathrm{K}_N^0 \supseteq \mathrm{P}_N \ltimes \mathrm{K}_1^{\times N}$.

For the non-interacting system, all of the single-particle symmetries and invariants still commute with the total Hamiltonian, including single-particle time translations. That is what the $N$-fold direct product of one-particle symmetries $\mathrm{K}_1^{\times N}$ encapsulates. The semidirect product $\ltimes$ captures that fact that particle permutations act as homomorphisms on the normal abelian subgroup $\mathrm{K}_1^{\times N}$, e.g.\ for the unitary representation $\hat{U}(ij)$ of the transposition $(ij)\in\mathrm{P}_N$, one finds that $\hat{U}(ij)\hat{H}_i = \hat{H}_j \hat{U}(ij)$. This symmetry is enough for every system to be integrable for any $N$ and $J$. A complete set of commuting observables for the spatial Hilbert space is  $\{\hat{H}_1^1, \hat{H}_2^1, \ldots\}$ and this is equivalent to diagonalizing the $\mathrm{K}_1^{\times N}$ subgroup. For symmetrizing identical particles and perturbation theory, diagonalizing the $\mathrm{P}_N$ subgroup is more useful. For $N=2$ and any trap, the system is superintegrable in a sense because $\hat{H}_0^2$ and $\hat{U}(12)$ are also sufficient to separate all of $\KHS$. For harmonic traps, $\hat{H}^N_0$ is maximally superintegrable because  $\mathrm{K}_N^0 \sim \mathrm{U}(N)$~\cite{Baker}.

\item Symmetry breaking from $\mathrm{K}_N^0$ to $\mathrm{K}_N$ for weak interactions is not algebraically universal for $N>4$.

Symmetry breaking is calculated using first-order perturbation theory aided by symmetrization of the subgroup  $\mathrm{P}_N$ subgroup common to both $\mathrm{K}_N^0$ and $\mathrm{K}_N$. For $N=2$, the magnitude of the first-order splitting energies  depends on the specific properties of $V^1$ and $V^2$, but the form is algebraically universal, i.e.\ the same algebraic equation for every trap and every two-body interaction, and not, for example, a set of transcendental equations. Further, the expressions for the first-order eigenstates does not depend on $V^1$ or $V^2$. For $N=3$ and $N=4$, there are still algebraically-universal expressions for the first-order energy shifts and corresponding eigenstates. However, some universality is lost because now both expressions depend on the trap and interaction for mixed-symmetry states. These states that are neither totally symmetric or totally antisymmetric are relevant when $J>1$. For $N>4$, there are no algebraically universal expressions for first-order perturbation theory because solutions in mixed symmetry subspaces require solving polynomial equations of order quintic or higher. There are of course specific solutions for specific potentials and traps, e.g.\ harmonic interaction in harmonic traps~\cite{Armstrong2012}.

\item Two-body matrix elements of the contact interaction potential have state permutation symmetry.

Consider the two-body matrix elements $\br{\alpha \beta}\hat{V}_{12}\kt{\gamma \zeta}$ of the two body interaction, where $\kt{\gamma \zeta} = \kt{\gamma} \otimes \kt{\zeta}$ is the tensor product of the two one-particle states satisfying $\hat{H}^1 \kt{\gamma} = \epsilon_\gamma \kt{\gamma}$. From Galilean invariance and hermiticity (and choosing phases so that the one-dimensional trap eigenstates are all real) the two-body matrix elements of the the contact interaction $V^2(|x_i - x_j|) = g \delta(x_i -x_j)$ have the property that all 24 permutations of the four states in $\br{\alpha \beta}\hat{V}_{12}\kt{\gamma \zeta}$ will have the same two-body matrix element. Since totally-antisymmetric spatial states are odd under permutations of state or particles, this symmetry explains why identical one-component fermions do not feel the contact interaction. For a flat potential with hard boundaries, i.e.\ the infinite square well, the two-body matrix elements are not only state permutation invariant, but they also do not depend any state property of $\kt{\alpha}$, $\kt{\beta}$, etc.\ except the for the degeneracy pattern of the quantum numbers in the the composition of the four states. The infinite square well with contact interactions is also integrable and solvable (but not algebraically solvable) for any value of $g$, number of particles $N$, and number of spin components $J$ using the Bethe ansatz~\cite{ansatz}.

\item In the unitary limit of contact interactions $g \rightarrow \infty$, ordering permutation symmetry emerges as a symmetry of the Hamiltonian $\hat{H}_\infty^N$.

The coincidence manifold $\mathcal{V}^N$ is the union of all the $(N-1)$-dimensional hyperplanes defined by $x_i = x_j$. The configuration space $\mathcal{X}^N$ is sectioned into $N!$ sectors $\mathcal{X}_s$ by the manifold $\mathcal{V}^N$, one for every permutation $s = \{s_1 s_2 s_3 \cdots s_N\} \in\mathrm{S}_N$.  Each sector $\mathcal{X}_s$ is the set of configurations with $x_{s_1} < x_{s_2} < \cdots < x_{s_N}$. In the unitary limit of the contact interaction, the manifold $\mathcal{V}^N$ is a nodal surface for all finite-energy states and there is no tunneling between sectors~\cite{bao}. Particle permutations $p \in\mathrm{P}_N$ act on $\mathcal{X}^N$ as orthogonal transformations and map sectors onto sectors like  $p \mathcal{X}_s = \mathcal{X}_{ s p^{-1}}$. Ordering permutations $\mathfrak{o} \in \mathfrak{O}_N$ rearrange the order of the positions, without regard to which particle is in which position. They are not linear transformations on $\mathcal{X}^N$ but they do map sectors onto sectors naturally: $\mathfrak{o} \mathcal{X}_s = \mathcal{X}_{\mathfrak{o} s}$. The set of ordering permutations $\mathfrak{O}_N$ is therefore also isomorphic to $\mathrm{S}_N$, but distinct from and commuting with $\mathrm{P}_N \sim \mathrm{S}_N$.

\item The unitary-limit Hamiltonian $\hat{H}_\infty^N$ always has at least the minimal kinematic symmetry  $\mathrm{K}_N^\infty \supseteq \mathrm{P}_N \times \mathfrak{O}_N \times \mathrm{K}_1$ when restricted to the subspace of finite energy states.

The finite-energy spectrum of $\hat{H}_\infty^N$ is the same as the spectrum of $\hat{H}_\infty^0$ restricted to totally antisymmetric spatial states. Therefore, for every distinct set of $N$ one-particle states, there is an energy level of $\hat{H}_\infty^N$ that is $N!$-fold degenerate before symmetrization. A complete set of commuting operators for $N$ distinguishable particles with spin $s$ ($J= 2s+1$) is provided by each particle's spin operator, e.g.\ $\hat{S}^z_i$, the total Hamiltonian $\hat{H}_\infty^N$, and a complete set of conjugacy class operators for canonical subgroup chains $\mathrm{P}_N \supset \mathrm{P}_{N_1} \supset \cdots \supset \mathrm{P}_2$ and $\mathfrak{O}_N \supset \mathfrak{O}_{N_1} \supset \cdots \supset \mathfrak{O}_2$. For indistinguishable fermions or bosons, these subgroup chains can be used (or in some cases modified to account for parity or spin composition) to select the state with the correct symmetrization properties~\cite{unitary}. The degeneracy of allowed states can be calculated algebraically using combinatorics. 

\item Symmetry breaking from $\mathrm{K}_N^\infty$ to $\mathrm{K}_N$ for near-unitary interactions is not algebraically universal for $N>4$.

In the near-unitary limit, $\mathfrak{O}_N$ symmetry is broken by tunneling between adjacent sectors but  $\mathrm{P}_N \times \mathrm{K}_1$ is preserved. The level-splitting can again be calculated by  degenerate perturbation theory~\cite{perturb}. This time, the irreducible representations of $\mathfrak{O}_N\sim \mathrm{S}_N$ must be diagonalized. For $N=2$ and $N=3$ in a symmetric trap, the specific splitting is proportional to $1/g$, and the form of the splitting is the same for every trap shape. For three particles in an asymmetric trap, four particles, and five particles in a symmetric trap, the specific splitting depends two or three different tunneling parameters, all proportional to $1/g$, but whose relative ratios depend on the trap shape. Therefore, there is less universality in this case, but the equations are still algebraically solvable and take the same form in terms of the tunneling parameters for every case. For five particles in an asymmetric trap or six or more particles, there cannot be algebraic universality for the same reason at is was lost for weak interactions: quintic equations (or worse) arise. Note that for all cases, the totally antisymmetric state is universal, and the splitting of the totally symmetric state is universal. The failure of algebraic universality is only relevant for mixed symmetry states.

\end{enumerate}

\section{Concluding Remarks}

One direction these comments on symmetry and integrability point is towards computational algorithms with increased efficiency. Another direction is implementation of quantum control protocol that exploits symmetry-protected subspaces of the Hamiltonian, perhaps for information processing. For example, if experimental control is sufficient, the symmetries of trapped ultracold gases in one-dimensional can be mapped on the symmetries of other quantum systems, enabling more robust simulation. And to conclude, if more is different, then five is more.

\section*{Acknowledgements}
Thanks to the organizers and participants of FB21 for a rewarding conference, especially colleagues in the Friday session on Low-Dimensional Systems for their stimulating presentations.

%
%
%

\end{document}